\documentclass{aa}
\usepackage[longnamesfirst]{natbib}
\usepackage{graphics}
\bibpunct{(}{)}{;}{a}{}{,}

\newcommand\msun{\ensuremath{M_\odot}}
\newcommand\rsun{\ensuremath{R_\odot}}
\newcommand\lsun{\ensuremath{L_\odot}}
\newcommand\rhos{\ensuremath{\sqrt{\bar{\rho}/\bar{\rho}_{\odot}}}}

\newcommand\phn{\ensuremath{\phantom{0}}}
\newcommand\phnn{\ensuremath{\phantom{00}}}

\newcommand\nodata{\ensuremath{~\cdots~}}
\newcommand\D{\ensuremath{\mathrm{d}}}
\newcommand\mc{\multicolumn}
\newcommand\heii{\ion{He}{ii}}
\newcommand\ec{\extracolsep}
\newcommand\ea{et al.}
\newcommand\jcd{Christensen-Dalsgaard}
\newcommand\aap{A\&A}
\newcommand\aapr{A\&AR}
\newcommand\aaps{A\&AS}
\newcommand\apj{ApJ}

\newcommand\mnras{MNRAS}

\newlength{\figwidth}
\newlength{\dblfigwidth}
\begin{document}

\title{Seismic study of stellar convective cores}
\titlerunning{Seismic study of stellar convective cores}
\author{Anwesh Mazumdar \and H. M. Antia}
\authorrunning{Mazumdar \and Antia}
\offprints{H. M. Antia}
\institute{Tata Institute of Fundamental Research,
Homi Bhabha Road, Mumbai 400005, India\\
email: anwesh@tifr.res.in, antia@tifr.res.in}
\date{Received \today}

\abstract{
It has been shown that a discontinuity in the derivatives of the sound speed
at the edge of the convective regions inside a star gives rise to a
characteristic oscillatory signal in the frequencies of stellar
oscillations. This oscillatory signal has been suggested as a means
to study the base of the outer convection zone in low mass stars and
possibly the outer edge of the convective core in high mass stars.
Using stellar models we show that because of a phenomenon similar
to aliasing in Fourier transform, it may not be possible to use this
signal to detect the convective core. Nevertheless, it may be possible to
determine the size of convective cores using the frequency
separation $\nu_{n+1,\ell}-\nu_{n,\ell}$.
}

\maketitle

\section{Introduction}
\label{sec:intro}

It has been shown that a discontinuity in the derivatives of the sound
speed in the stellar interior, like those arising at the edge of a
convective region, introduces a characteristic oscillatory signature in
the frequencies of low degree modes \citep{gough:90,mct:94,rv:94,ban:94}.
From this oscillatory signature it has been possible to put limits on the
extent of overshoot below the solar convection zone \citep{basu:97}.
\citet{mct:00} have pointed out that this oscillatory signature can be
used to study the location of the base of the convection zone as well as
the extent of overshoot below this base in other stars using asteroseismic
data for only low degree modes. \citet{mct:98} have also suggested that
the same oscillatory signal can be used to study the size of convective
cores in massive stars. However, it is not clear if such a signal has been
demonstrated using actual stellar models with convective cores.

Similarly, helioseismic inversions for the rotation rate in the solar
interior \citep{thompson:96,schou:98} have shown the presence of a
tachocline \citep{sz:92} around the base of the convection zone, where the
rotation rate varies from differential rotation inside the convection zone
to almost solid body like rotation in the radiative interior. This sharp
transition in the rotation rate introduces an oscillatory signature in the
frequency splitting coefficients \citep{ma:01}, which can be used to study
the characteristics of tachoclines in other stars.

In this work we attempt to study the possible oscillatory signal arising
from convective cores in stars more massive than the Sun. However, we do
not find any significant oscillatory signature of the core and try to
identify the reason for the apparent absence of this signal. We also study
how the amplitude of oscillatory signal from the outer convection zone
varies with stellar mass and age. Since even the signal from the outer
convection zone is rather weak, such studies may help us in identifying
promising stars where attempts can be made to detect such an oscillatory
signal. In the absence of a substantial oscillatory signal from the
convective core of stars, we try to identify alternate means to detect the
presence of a convective core and to measure its size.

\section{The technique}
\label{sec:technique}

In the asymptotic limit \citep{cb:91} the eigenfunctions of stellar
oscillation modes have the general form $\cos(\omega\tau+\phi)$, where
$\tau$ is the acoustic depth given by
\begin{equation}
\tau=\int_r^R {1 \over c(r)}\; \D r.
\label{eq:tau}
\end{equation}
Here $R$ is the stellar radius. Using this form for the eigenfunctions it
can be shown \citep{mct:94} that if we have a discontinuity in the
derivatives of the sound speed at a point corresponding to an acoustic
depth $\tau_c$ then the frequencies of p-modes can be written as
\begin{equation}
\omega_{n,\ell}=\omega^{(s)}_{n,\ell}+
A(\omega_{n,\ell})\sin(2\omega_{n,\ell}\tau_c+\phi),
\label{eq:split_form}
\end{equation}
where $n$ is the radial order and $\ell$ the degree of the mode and
$\omega^{(s)}_{n,\ell}$ is the smooth part of the frequency arising from
the smooth variation in the sound speed with depth, while the second term
is the oscillatory component. The amplitude $A(\omega)$ is a smooth
function of $\omega$. Here we have neglected terms involving the degree
$\ell$ since their effects are small for low degree ($\ell\le3$) modes. In
this work we are primarily interested in stars other than the Sun, where it
is not possible to detect oscillations with higher degree.  Similar
oscillatory terms can arise in the splitting coefficients if there is any
steep variation in the rotation rate such as what happens in the
tachocline region \citep{ma:01}.

Following \citet{ban:94}, we take the fourth difference of the frequencies
with respect to $n$ to enhance the oscillatory signal. Another advantage
of taking the fourth difference is that the smooth part of the frequencies
becomes negligible and we do not need to include it in our analysis. This
will of course, depend on the smooth component of variation of the sound
speed, but at least for the Sun this component is found to be negligible.
The sharp variation in sound speed inside the second helium ionisation
zone also gives rise to an oscillatory signal, but in this case the
region with reduced adiabatic index, $\Gamma_1$, has a small but finite 
width and in
some sense it can be considered as two steps of opposite sign located
close to each other.
\citet{mt:98} have shown that the amplitude of such a signal
would be modulated with an oscillatory factor of the form
$\sin^2(\omega\beta)$, where $\beta$ is the acoustic half-width of
the \heii\ ionisation zone. We have not included this factor
in our adopted form for the oscillatory signal.
Thus for the Sun there are two oscillatory components -- one arising from the
base of the convection zone and the other from the second helium
ionisation zone. A similar form may be expected for other stars with $M\la
1.2M_\odot$, where there is no convection in the core. For more massive
stars with convective cores we may expect a third component arising from
the boundary of the convective core.

Since there are no observations yet of stellar oscillations with
sufficient accuracy to detect this oscillatory signal, in this work we
construct stellar models with different mass and age and calculate their
oscillation frequencies. Using the computed frequencies we attempt to
check if the location of these regions of rapid variation in sound speed
can be determined using the frequencies of only low degree modes, which
are likely to be observed by forthcoming asteroseismic missions. 

Apart from frequencies, we also study the oscillatory signal in the
splitting coefficients arising from possible tachoclines in stars. There
is no theory to determine the position of tachoclines in stars, but for
simplicity we generally assume that the tachocline is located at the base
of the outer convection zone, which is in fact the case for the Sun
\citep{sz:92}. However, in some cases we have studied the effect of
varying the position of the tachocline to see how the oscillatory signal
varies with its depth. The main advantage of using the splitting
coefficients arising from an assumed tachocline is that, in this case
there is only one layer where the rotation rate changes rapidly and its
location can be altered to study how the signal varies. This exercise is
not possible with frequencies where there are multiple layers with rapid
variation in the sound speed and their locations cannot be easily shifted
over the entire stellar radius.

In order to study the oscillatory signature in splitting coefficients
we assume a model tachocline rotation profile of the form
\citep{ma:01}
\begin{equation}
\Omega(r,\theta)={\delta\Omega (5\cos^2\theta -1)\over 1+\exp[(r_d-r)/w]},
\label{eq:rotprof}
\end{equation}
where $\delta\Omega$ is the extent of variation in the rotation rate
across the tachocline, lying at a radial position of $r_d$ and having a
half-width of $w$. Using this model rotation profile we can calculate the
corresponding splitting coefficients for a star. Because of the choice of
latitudinal dependence only the splitting coefficient $c_3(n,\ell)$ is
found to be non-zero. Neglecting the smooth part, the fourth difference
can be written as
\begin{equation}
\delta^4 c_3(n,\ell)
   =\left(a_0+{a_1\over \omega_{n,\ell}}+{a_2\over\omega_{n,\ell}^2}\right)
\sin(2\omega_{n,\ell}\tau+\phi).
\label{eq:fourth_fit}
\end{equation}
The parameters $a_0,a_1,a_2,\tau$ and $\phi$ can be determined by a
nonlinear least squares fit. For the frequencies also we use the same form
for fitting except that there are multiple oscillatory terms. Each
oscillatory part is defined by a similar set of parameters, but the number
of parameters to be fitted is larger and hence it is more difficult to
obtain a proper fit. For distant stars it will be possible to detect modes
with $\ell=0,1,2,3$ only and hence we use only these modes in our study.
As mentioned earlier, the amplitude of the signal arising from
the \heii\ ionisation zone has an oscillatory factor and may not be
properly represented by the form we have adopted. 
However, the half-width $\beta$ being quite small, the `wavelength' of
oscillation in amplitude would be quite large and in the limited
frequency range that is likely to be observed we expect less than
half a wavelength to be accommodated. Such a variation can be
approximated reasonably well by the three terms that we have
incorporated. Hence we have little difficulty in fitting the
oscillatory signal from the \heii\ ionisation zone.

\begin{figure}
\centering
\resizebox{\figwidth}{!}{\includegraphics{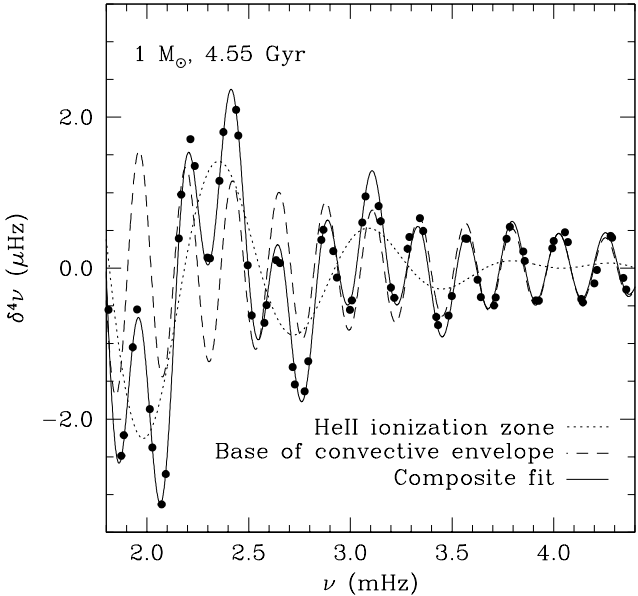}}
\caption{
	Fourth difference of the frequency $\nu$, for a solar mass star at
	an age of $4.55$~Gyr. A fit to the data using a function having
	two oscillatory components of the form described in
	Eq.~(\ref{eq:fourth_fit}) is shown. Each oscillatory component is
	also shown separately in the figure. 
}
\label{fig:osc_fit}
\end{figure}
To illustrate the oscillatory signal, Fig.~\ref{fig:osc_fit} shows a fit
to the fourth difference of frequencies for a star with $M= 1~M_\odot$ and
age $4.55\times 10^9$ years. This model is slightly different from the
current solar models, since diffusion of helium and heavy elements in the
radiative interior is neglected. One can see the two oscillatory
components in this figure. The slowly varying component with $\tau\approx
676$~s arises from the second helium ionisation zone and the value of
$\tau$ is approximately equal to the acoustic depth of this ionisation
layer. The more rapidly oscillating component ($\tau \approx 2176$~s)
arises from the base of the convection zone. The amplitudes of all
oscillatory components decrease with frequency. At still higher
frequencies where the modes are not trapped below the solar surface the
effective value of $\tau$ for the fits may also change. We do not include
such modes as they are not likely to be observed and in any case their
interpretation is difficult. Thus for all stellar models we generally
restrict ourselves to frequencies less than the acoustic cutoff frequency
$\omega_{ac} \approx {c\over 2H_p}$ where $c$ is the sound speed and $H_p$
is the pressure scale height near the stellar surface. At low frequencies
the fourth difference generally shows steep variations, probably because
the low radial order modes do not conform to the asymptotic relations used
in obtaining the oscillatory form. Thus we neglect these modes too while
studying the oscillatory signal. The frequency range studied for different
stars varies with mass and age and in order to compare the respective
amplitudes of the oscillatory signal we scale the frequencies by
$\sqrt{M/R^3}$, where $M$ is the stellar mass and $R$ the radius in terms
of solar values. Thus we compare the amplitudes of oscillatory signal at
frequency $\nu_0\rhos$, where $\nu_0$ is a fixed value chosen for the Sun
and $\bar\rho$ is the mean density.

\section{The stellar models}
\label{sec:models}

In order to study the variation in the oscillatory signal for different
stars we construct models for stars with different masses at various ages
using the stellar evolution code CESAM \citep{morel:97}. For simplicity we
have neglected diffusion of helium and heavy elements in stellar interior.
We use the EFF \citep{eff:73} equation of state with Coulomb correction
\citep{cd:92}; the OPAL \citep{ir:96} opacity tables; nuclear reaction
rates from the NACRE compilation \citep{aar:99,morel:99}; mixing length of
$1.8 H_p$, where $H_p$ is the pressure scale height.

We have considered stellar models with
$M/M_\odot=0.5,0.75,1,1.25,1.5,1.75,2,2.5,3,4$ and $5$ at different ages
spanning the main sequence phase of evolution. The ZAMS chemical
composition of all the models were taken to be solar-like ($X = 0.708, Z =
0.019$). The characteristics of some of the representative models are
summarised in Table~\ref{tab:models}. Here, $X_c$ is the hydrogen
abundance at the centre, $\nu_{ac}=c/(4\pi H_p)$ is the acoustic cutoff
frequency at the surface, $\tau_0$ is the acoustic radius of the star or
the sound travel time from the centre to the surface. The radial distance
as well as the acoustic depth of the outer edge of the convective core,
base of the convective envelope and the \heii\ ionisation zone are also
given in the table.

\begin{table*}
\centering
\caption{
	Characteristics of stellar models considered
}
\label{tab:models}
\begin{tabular}{ccccccc@{\ec{8pt}}c@{\ec{0pt}}c@{\ec{8pt}}c@{\ec{0pt}}c@{\ec{8pt}}c@{\ec{0pt}}c}
\hline
$M/\msun$ & Age & $R/\rsun$ & $L/\lsun$ &
$X_c$ & $\nu_{\scriptscriptstyle\mathrm{ac}}$ & $\tau_0$ &
\mc{2}{c}{Convective core} &
\mc{2}{c}{Convective env.} &
\mc{2}{c}{\heii\ ionisation} \\
\cline{8-9} \cline{10-11} \cline{12-13}
& & & & & & &
$r_{\scriptscriptstyle\mathrm{core}}/R$ & 
$\tau_{\scriptscriptstyle\mathrm{core}}$ &
$r_{\scriptscriptstyle\mathrm{env}}/R$ &
$\tau_{\scriptscriptstyle\mathrm{env}}$ &
$r_{\scriptscriptstyle\mathrm{\heii}}/R$ & 
$\tau_{\scriptscriptstyle\mathrm{\heii}}$ \\
 & (Gyr) & & & & (mHz) & (s) & & (s) & & (s) & & (s) \\
\hline
0.50 & 0.00 & 0.460 & \phnn 0.034 & 0.708 &     15.092 &  1609 &\nodata& \nodata   & 0.632 &     1060 & 0.977 & \phn 320 \\
0.50 & 2.00 & 0.479 & \phnn 0.033 & 0.692 &     14.093 &  1711 &\nodata& \nodata   & 0.614 &     1150 & 0.976 & \phn 350 \\
0.50 & 8.00 & 0.484 & \phnn 0.034 & 0.649 &     13.759 &  1739 &\nodata& \nodata   & 0.611 &     1171 & 0.976 & \phn 355 \\
\hline                                                                                           
0.75 & 0.00 & 0.640 & \phnn 0.204 & 0.708 &     10.218 &  2139 & 0.115 & \phn 2039 & 0.687 &     1327 & 0.981 & \phn 396 \\
0.75 & 2.00 & 0.666 & \phnn 0.198 & 0.654 & \phn 9.560 &  2276 &\nodata& \nodata   & 0.670 &     1443 & 0.980 & \phn 429 \\
0.75 & 8.00 & 0.689 & \phnn 0.230 & 0.498 & \phn 8.837 &  2394 &\nodata& \nodata   & 0.664 &     1530 & 0.980 & \phn 459 \\
\hline                                                                                           
1.00 & 0.00 & 0.861 & \phnn 0.694 & 0.708 & \phn 6.929 &  2885 & 0.147 & \phn 2722 & 0.724 &     1700 & 0.982 & \phn 516 \\
1.00 & 4.55 & 0.976 & \phnn 0.996 & 0.357 & \phn 5.321 &  3464 &\nodata& \nodata   & 0.721 &     2062 & 0.980 & \phn 645 \\
1.00 & 7.00 & 1.071 & \phnn 1.241 & 0.136 & \phn 4.399 &  3965 &\nodata& \nodata   & 0.715 &     2393 & 0.979 & \phn 763 \\
\hline                                                                                           
1.25 & 0.00 & 1.159 & \phnn 1.756 & 0.708 & \phn 4.585 &  4026 & 0.140 & \phn 3823 & 0.789 &     2097 & 0.983 & \phn 684 \\
1.25 & 2.00 & 1.325 & \phnn 2.731 & 0.374 & \phn 3.417 &  4928 & 0.053 & \phn 4839 & 0.847 &     2198 & 0.983 & \phn 797 \\
1.25 & 3.50 & 1.531 & \phnn 3.116 & 0.069 & \phn 2.616 &  6045 & 0.053 & \phn 5929 & 0.791 &     3161 & 0.979 &     1104 \\
\hline                                                                                           
1.50 & 0.00 & 1.504 & \phnn 3.777 & 0.708 & \phn 3.149 &  5498 & 0.129 & \phn 5262 & 0.881 &     2133 & 0.985 & \phn 802 \\
1.50 & 1.00 & 1.640 & \phnn 5.786 & 0.457 & \phn 2.529 &  6438 & 0.078 & \phn 6285 & 0.958 &     1352 & 0.990 & \phn 641 \\
1.50 & 1.90 & 2.004 & \phnn 6.340 & 0.115 & \phn 1.786 &  8354 & 0.056 & \phn 8198 & 0.874 &     3336 & 0.982 &     1291 \\
\hline                                                          
1.75 & 0.00 & 1.829 & \phnn 7.191 & 0.708 & \phn 2.372 &  7061 & 0.122 & \phn 6794 & 0.964 &     1367 & 0.991 & \phn 665 \\
1.75 & 0.90 & 2.064 & \phn 11.451 & 0.331 & \phn 1.763 &  8566 & 0.075 & \phn 8375 & 0.991 & \phn 658 & 0.993 & \phn 552 \\
\hline                                                         
2.00 & 0.00 & 2.015 & \phn 12.505 & 0.708 & \phn 2.042 &  7830 & 0.125 & \phn 7534 & 0.992 & \phn 562 & 0.994 & \phn 485 \\
2.00 & 0.50 & 2.064 & \phn 19.045 & 0.427 & \phn 1.720 &  8127 & 0.091 & \phn 7909 & 0.991 & \phn 628 & 0.993 & \phn 546 \\
\hline                                                         
2.50 & 0.00 & 2.296 & \phn 31.209 & 0.708 & \phn 1.579 &  8609 & 0.134 & \phn 8257 & 0.991 & \phn 634 & 0.993 & \phn 552 \\
2.50 & 0.30 & 2.452 & \phn 47.999 & 0.395 & \phn 1.249 &  9454 & 0.094 & \phn 9189 & 0.991 & \phn 717 & 0.992 & \phn 623 \\
\hline                                                         
3.00 & 0.00 & 2.546 & \phn 65.099 & 0.708 & \phn 1.301 &  9211 & 0.143 & \phn 8803 & 0.992 & \phn 637 & 0.993 & \phn 549 \\
3.00 & 0.20 & 2.831 &     101.563 & 0.364 & \phn 0.930 & 10707 & 0.095 &     10401 & 0.991 & \phn 746 & 0.993 & \phn 642 \\
\hline                                                         
4.00 & 0.00 & 2.983 &     201.651 & 0.708 & \phn 0.904 & 10130 & 0.160 & \phn 9616 & 0.993 & \phn 569 & 0.995 & \phn 475 \\
4.00 & 0.08 & 3.079 &     295.115 & 0.445 & \phn 0.825 & 10563 & 0.119 &     10171 & 0.994 & \phn 561 & 0.995 & \phn 460 \\
\hline                                                         
5.00 & 0.00 & 3.365 &     470.183 & 0.708 & \phn 0.853 & 10873 & 0.175 &     10255 & 0.995 & \phn 501 & 0.996 & \phn 405 \\
5.00 & 0.05 & 3.584 &     695.338 & 0.420 & \phn 0.725 & 11884 & 0.123 &     11421 & 0.995 & \phn 535 & 0.996 & \phn 415 \\
\hline

\end{tabular}
\end{table*}

For each of these models we calculate the frequencies of p-modes for
$\ell=0,1,2,3$ and use them to study the oscillatory signal using the
second or fourth difference. The oscillatory signal also depends on the
extent of overshoot below the convection zone \citep{mct:94} as well as the
treatment for calculating the diffusion of helium and heavy elements in
the radiative regions \citep{ba:94}. From the solar studies it has been
known that the overshoot below the base of the convection zone is quite
small \citep{basu:97} and further the amplitude of the observed signal is
consistent with that in a solar model without overshoot and with very
little composition gradient at the base of the convection zone. The
absence of composition gradient is probably due to some mixing process
operating in that region \citep{richard:96,btz:99}. Thus in this work we
neglect both diffusion and overshoot. Of course, it is possible that in
some stars the overshoot may be larger, but that will only increase the
amplitude of the oscillatory signal. Similarly, a gradient in helium
abundance below the outer convection zone also tends to enhance the
oscillatory signal. Thus by neglecting overshoot and diffusion we will get
a lower limit on the amplitude of the oscillatory signal. For massive
stars where the outer convection zone becomes shallow, it is quite likely
that extent of overshoot as well as composition gradient may be larger
than that in the Sun, because the efficiency of convection would be lower
\citep{saikia:00}.

Apart from this, for most of these models we also calculate the splitting
coefficients assuming a model tachocline profile given by
Eq.~(\ref{eq:rotprof}). We use $\delta\Omega=20$~nHz (which is the typical
value for the Sun) and $w=0.001R$ while $r_d$ is taken to be the base of
the outer convection zone. There is no established theory to suggest that
the tachocline has to coincide with the base of the convection zone, but
it is quite likely to be true. In some cases, we study the variation in
the oscillatory signal with the depth of the tachocline. The width is
chosen to be somewhat small as in massive stars the tachocline may be
located in regions where the scale heights are smaller than that for the
solar case. The effect of the width of the tachocline on the oscillatory
signal has been discussed by \citet{ma:01}. For stars with $M\ga
1.5M_\odot$ the outer convection zone is too shallow and even a width of
$0.001R$ may be too large. But such stars are unlikely to have tachoclines
and we do not consider this signal in those cases. The variation of
amplitude in the splitting coefficients with stellar mass and age is
similar to that in the frequencies. We still consider the two separately
as the splitting coefficients provide a convenient set to study the
variation in oscillatory signal with depth, which in this case can be
easily controlled.

\section{Signal from the base of outer convection zone}
\label{sec:czbase}

In order to study the oscillatory signature in the frequencies we
calculate the frequencies for each stellar model and fit the fourth
differences of frequencies to an oscillatory signal described in
Eq.~(\ref{eq:fourth_fit}), with two oscillatory components of the same
form. The results are shown in Table~\ref{tab:fit_comp}. This table also
includes the averaged frequency separations
$\Delta\nu_0=\nu_{n+1,0}-\nu_{n,0}$, $D_0=(\nu_{n,0}- \nu_{n-1,2})/6$, and
$d_{1/2} = (3/2)(\nu_{n,0} - 2\nu_{n,1} + \nu_{n+1,0})$. The fitted $\tau$
for the two components are close to the actual acoustic depths of the \heii\ 
ionisation zone and the base of the convection zone. Thus it is clear that
the depths of these layers can be estimated from the measured frequencies
if the accuracy is sufficient to fit the oscillatory signal.

\begin{table*}
\centering
\caption{
	Results from fits to oscillatory signal
}
\label{tab:fit_comp}
\begin{tabular}{c@{\ec{3pt}}c@{\ec{5pt}}c@{\ec{3pt}}c@{\ec{10pt}}c@{\ec{8pt}}c@{\ec{5pt}}c@{\ec{10pt}}c@{\ec{8pt}}c@{\ec{12pt}}c@{\ec{8pt}}c@{\ec{12pt}}c@{\ec{10pt}}c@{\ec{8pt}}c@{\ec{12pt}}c}
\hline
& & & &
\mc{3}{c}{Frequency separations} &
\mc{5}{c}{Convective envelope} &
\mc{3}{c}{\heii\ ionisation} \\
\cline{5-7} \cline{8-12} \cline{13-15}
$M/\msun$ & Age & $X_c$ & $\rhos$ &
$\Delta \nu_{0}$ & $D_{0}$ & $d_{1/2}$ &
Model &
\mc{2}{c}{Sp.\ coeff.\ fit} &
\mc{2}{c}{Frequency fit} &
Model &
\mc{2}{c}{Frequency fit} \\
\cline{9-10} \cline{11-12} \cline{14-15}
& & & & & & & 
$\tau^{\scriptscriptstyle\mathrm{(m)}}_{\scriptscriptstyle\mathrm{env}}$ &
$\tau^{\scriptscriptstyle\mathrm{(s)}}_{\scriptscriptstyle\mathrm{env}}$ &
$A^{\scriptscriptstyle\mathrm{(s)}}_{\scriptscriptstyle\mathrm{env}}$ &
$\tau^{\scriptscriptstyle\mathrm{(f)}}_{\scriptscriptstyle\mathrm{env}}$ &
$A^{\scriptscriptstyle\mathrm{(f)}}_{\scriptscriptstyle\mathrm{env}}$ &
$\tau^{\scriptscriptstyle\mathrm{(m)}}_{\scriptscriptstyle\mathrm{\heii}}$ &
$\tau^{\scriptscriptstyle\mathrm{(f)}}_{\scriptscriptstyle\mathrm{\heii}}$ &
$A^{\scriptscriptstyle\mathrm{(f)}}_{\scriptscriptstyle\mathrm{\heii}}$ \\
& (Gyr) & & & ($\mu$Hz) & ($\mu$Hz) & ($\mu$Hz) & 
(s) & (s) & (nHz) & 
(s) & ($\mu$Hz) &
(s) &
(s) & ($\mu$Hz) \\
\hline
0.50 & 0.00 & 0.708 & 2.264 &     307.6 & 3.124 &     18.587 & 1060 & 1093 & 0.114 & 1095 & 0.096 & \phn 320 & \phn 315 & 0.037 \\
0.50 & 2.00 & 0.693 & 2.135 &     289.6 & 2.833 &     16.685 & 1150 & 1183 & 0.114 & 1185 & 0.082 & \phn 350 & \phn 354 & 0.015 \\
0.50 & 8.00 & 0.649 & 2.102 &     284.9 & 2.642 &     15.580 & 1171 & 1206 & 0.112 & 1211 & 0.078 & \phn 355 & \phn 350 & 0.033 \\
\hline                                                                                                 
0.75 & 0.00 & 0.708 & 1.693 &     229.5 & 2.926 &     17.516 & 1327 & 1386 & 0.112 & 1389 & 0.088 & \phn 396 & \phn 386 & 0.261 \\
0.75 & 2.00 & 0.654 & 1.594 &     215.8 & 2.420 &     14.493 & 1443 & 1505 & 0.111 & 1509 & 0.081 & \phn 429 & \phn 416 & 0.206 \\
0.75 & 8.00 & 0.498 & 1.514 &     205.3 & 1.951 &     11.463 & 1530 & 1597 & 0.110 & 1601 & 0.078 & \phn 459 & \phn 444 & 0.203 \\
\hline                                                                                                 
1.00 & 0.00 & 0.708 & 1.251 &     170.0 & 2.512 &     15.015 & 1700 & 1791 & 0.117 & 1789 & 0.069 & \phn 516 & \phn 531 & 0.436 \\
1.00 & 4.55 & 0.357 & 1.037 &     142.1 & 1.422 & \phn 8.336 & 2062 & 2176 & 0.115 & 2181 & 0.056 & \phn 645 & \phn 676 & 0.369 \\
1.00 & 7.00 & 0.136 & 0.902 &     124.2 & 0.845 & \phn 5.948 & 2393 & 2521 & 0.115 & 2524 & 0.045 & \phn 763 & \phn 785 & 0.301 \\
\hline                                                                                             
1.25 & 0.00 & 0.708 & 0.896 &     122.3 & 2.066 &     12.577 & 2097 & 2230 & 0.121 & 2237 & 0.047 & \phn 684 & \phn 747 & 0.585 \\
1.25 & 2.00 & 0.374 & 0.733 &     100.4 & 1.190 & \phn 6.780 & 2198 & 2315 & 0.125 & 2336 & 0.048 & \phn 797 & \phn 873 & 0.574 \\
1.25 & 3.50 & 0.069 & 0.590 & \phn 81.7 & 0.652 & \phn 8.481 & 3161 & 3332 & 0.124 & 3338 & 0.039 &     1104 &     1114 & 0.260 \\
\hline

\end{tabular}
\end{table*}

It can be easily shown that taking the second difference of frequencies
will enhance the oscillatory signal by a factor of
$4\sin^2(2\pi\tau\Delta\nu)$, where $\Delta\nu$ is the frequency
separation $(\nu_{n+1,\ell}-\nu_{n,\ell})$. For $\tau\Delta\nu<1/12$ or
$\tau\Delta\nu>5/12$ this factor will be smaller than unity and the
amplitude of the oscillatory term will reduce when higher order
differences are taken. Thus in such cases it may be better to use the
second difference or to fit the oscillatory term to the frequencies
themselves. In the latter case one will need some model to fit the smooth
part also. It can be easily seen that if the smooth part is assumed to be
linear in $n$, then it will vanish when second difference is taken. Thus
we can fit the second difference without worrying about the smooth part.
Similarly, if the smooth part is assumed to be a cubic polynomial in $n$,
it will vanish if the fourth difference is taken. For more general smooth
functions, e.g., splines, the smooth part will not necessarily vanish when
differences are taken but it will generally become smaller. However, the
errors also increase as higher order differences are computed. Thus,
fourth differences may be used only if $\sin^2(2\pi\tau\Delta\nu)>
1/\sqrt{2}$ \citep{ban:94}.
\begin{figure*}
\centering
\hbox to \hsize{\hfil
\resizebox{\dblfigwidth}{!}{\includegraphics{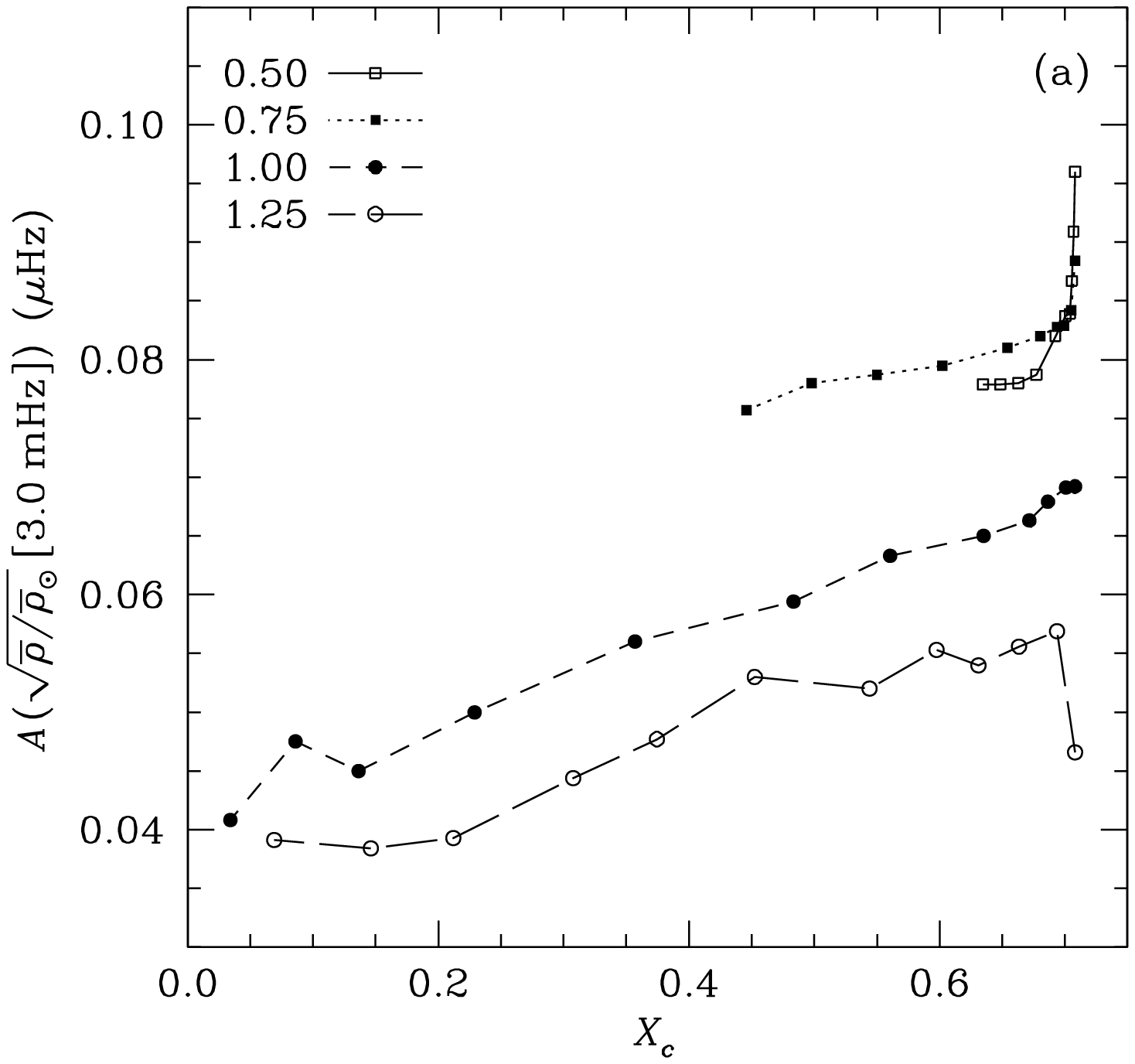}}
\hfil
\resizebox{\dblfigwidth}{!}{\includegraphics{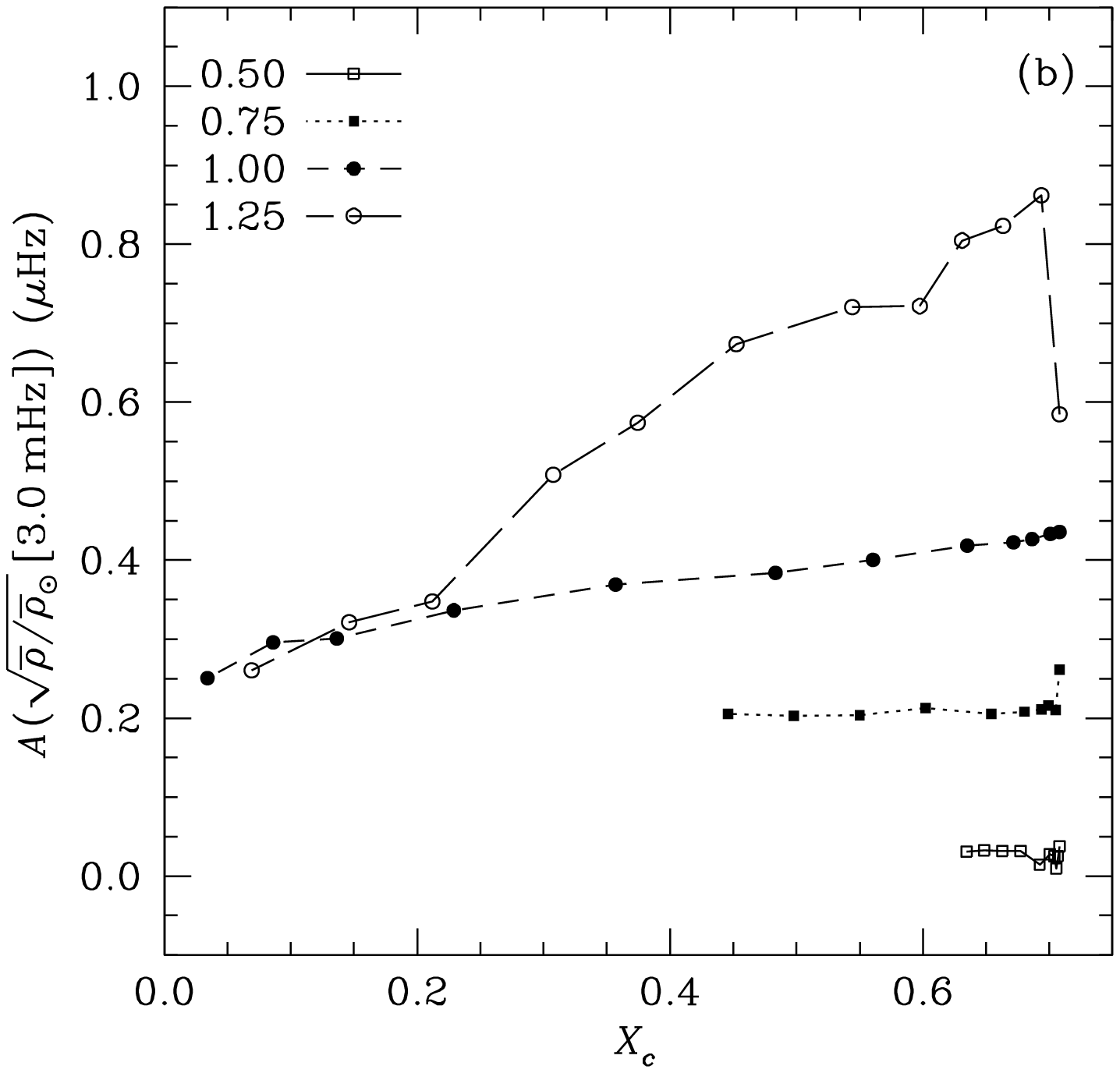}}
\hfil}
\caption{
	Comparison of amplitude of oscillatory signal in the frequency for
	evolving stellar models of different masses as marked in the
	figure. All the amplitudes are
	calculated at a scaled frequency corresponding to $3$~mHz for the
	present Sun. {\bf a)} The amplitude of the signal arising from the
	base of the convective envelope is shown. {\bf b)} The amplitude of
	the signal arising from the \heii\ ionisation region is shown.
	}
\label{fig:ampl_freq}
\end{figure*}

\begin{figure}
\centering
\resizebox{\figwidth}{!}{\includegraphics{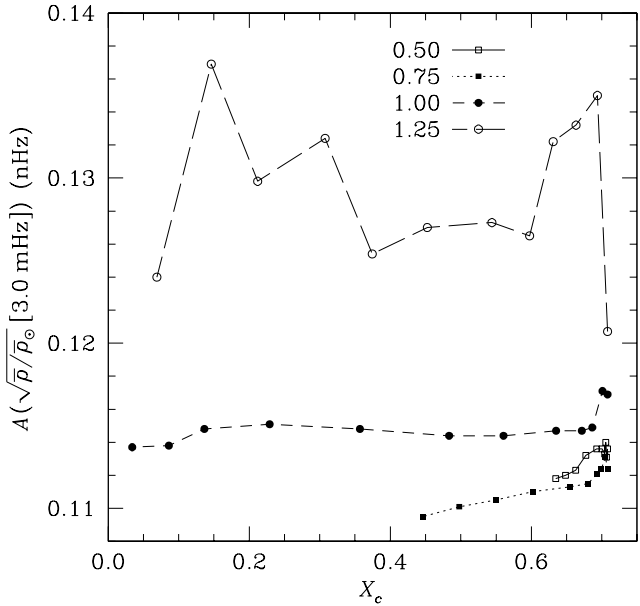}}
\caption{
	Comparison of amplitude of oscillatory signal in the splitting
	coefficients for evolving stellar models of different masses,
	arising due to a tachocline at the base of the convective envelope. 
	All the amplitudes are calculated at a scaled frequency corresponding
	to $3$~mHz for the present Sun. Note the difference in y-axis
	range between this figure and Fig.~\ref{fig:ampl_freq}.
	}
\label{fig:ampl_split}
\end{figure}

If $\tau_0$ is the acoustic radius or the sound travel time from the
centre to the surface of the star, then $\Delta\nu\approx 1/(2\tau_0)$.
Thus the amplification factor for second difference is
$4\sin^2(\pi\tau/\tau_0)$, which is maximum when $\tau\approx\tau_0/2$ or
when the layer is halfway between the centre and the surface in terms of
sound travel time. On the other hand, if the layer is close to either the
surface or the centre, the amplification factor will be small. Thus for
massive stars where the outer convection zone is shallow and the
convective core is also small in terms of acoustic radius, the fourth
difference will reduce the amplitude significantly. In such cases it may
be better to use alternate strategy to fit oscillatory signal. In this
work since we are using exact frequencies from stellar models the errors
are not amplified and hence as long as the amplitude does not decrease
significantly when differences are taken we can use the differences to
conveniently suppress the smooth part and isolate the oscillatory signal.
Even without
errors we find that when the outer convection zone becomes too shallow the
oscillatory signal reduces significantly when fourth differences are
taken. In such cases we have instead fitted the oscillatory signal in
second differences. Thus in order to facilitate comparison and remove the
variation in amplitude arising from variation of the amplifying factor, we
reduce all amplitudes to those in frequencies (or splitting coefficients)
by accounting for the amplification factor in taking second or fourth
differences. Thus the amplitudes given in Table~\ref{tab:fit_comp} or
those shown in Figs.~\ref{fig:ampl_freq}, \ref{fig:ampl_split} refer to
the frequencies or splitting coefficients, irrespective of whether second
or fourth differences were used in the fits. Another advantage of scaling
the amplitudes to remove the amplification factor is that variation in
the resulting amplitudes is significantly reduced.

Theoretically, there is no reason to believe that taking fourth differences
of frequencies will remove the smooth component. But in practice, using
a wide range of stellar models, we find that taking the
fourth differences essentially removes the smooth component of the
frequency in all cases, thus making it much simpler to fit the 
resulting oscillatory signal. 
While fitting the second difference or the frequencies directly, it
is essential to account for the smooth part, which necessarily involves
some assumptions about how the smooth component is represented. The resulting
fit will also depend to some extent on these assumptions.
In general, the smooth part
also depends on $\ell$ which makes it more difficult to remove.
For more rapidly varying
signal with $\tau\approx\tau_0/2$ like that arising from the base of the solar
convection zone, it may be relatively easy to characterise the smooth
part as has been done by \citet{mct:94}.
But when the
`wavelength' of oscillation is comparable to the frequency range in
which observations are available, it is not easy to
distinguish between the smooth and oscillatory components.
Taking the differences also tends to reduce the amplitude in such cases making
it more difficult to fit the oscillatory signal. We believe that this
difficulty is more fundamental, as it arises from the fact that there
is little difference between the smooth component and an
oscillatory component
with large wavelength. This problem will manifest itself in
some form or other irrespective of whether we fit the differences or the
frequencies directly. If taking the fourth differences wipes out the
oscillatory signal, then it would be possible to approximate it well
by a cubic polynomial, which can define the smooth part.

Fig.~\ref{fig:ampl_freq} shows the variation in amplitude of the fitted
signal in the frequencies with stellar age for a few stellar masses. It
can be seen that at masses $<1.5 M_\odot$ there is not much variation in
amplitude with age and further even the variation with stellar mass is
also quite small in the signal from the base of the convective envelope.
The maximum difference being about a factor of two between all models
included in the figure.
The oscillatory signal from the \heii\ ionisation zone has in general larger
amplitude and the amplitude increases with stellar mass. For $M>M_\odot$
the amplitude of signal from \heii\ ionisation zone decreases with ages,
while at lower mass it is essentially independent of age.  For more
massive stars ($M\ga 1.5M_\odot$) the acoustic depth of both the \heii\ 
ionisation zone and the base of convective envelope are comparable and it
becomes difficult to fit the two signals separately. In addition, because
the resulting $\tau$ is quite small, taking the differences tend to reduce
the amplitudes and it becomes more difficult to fit the signal even in the
second differences. In some cases, when the frequencies of two components
are close the resulting pattern shows beats in the limited frequency range
that is likely to be available. The apparent amplitude of the signal
depends on the relative phase of the two components and it is difficult to
obtain any reliable fit to both the components. The characteristics of the
signal arising from the base of the convective envelope may also be
different in these stars as the outer convection zone splits into two
parts. Consequently, we have not shown the results for these stars.

Apart from the frequencies we also calculate the splitting coefficients
for each of the stellar models with a model rotation profile given by
Eq.~(\ref{eq:rotprof}). This rotation profile assumes a tachocline at the
base of the outer convection zone. The fourth difference of the calculated
splitting coefficient $c_3(n,\ell)$ for $\ell=2,3$ is again fitted to an
oscillatory signature of the form given in Eq.~(\ref{eq:fourth_fit}) to
calculate the characteristics of the tachocline. These results are also
included in Table~\ref{tab:fit_comp} and it is clear that the fitted
$\tau$ is close to the actual depth of the tachocline in the model
profile. The variation of amplitude with stellar mass and age is shown in
Fig.~\ref{fig:ampl_split}. In this case, the variation with stellar
mass and age is very small.

The typical amplitudes of the oscillatory signal in the frequencies due to
the base of outer convection zone is $0.05~\mu$Hz, while that due to \heii\ 
ionisation zone is $0.5~\mu$Hz. We have not added any errors in the model
frequencies and hence there is little difficulty in identifying this
signal for low mass ($M< 1.5 M_\odot$) stars. Actual observations will
naturally have some random error associated with each frequency. From
simulations with artificial data sets it turns out that it is possible to
detect the oscillatory signal as long as the errors in frequencies are
less than or of the order of the amplitude of oscillations. Thus we need
an accuracy of about $0.05~\mu$Hz to detect this signal from the base of
convective envelope and measure its characteristics. For $\tau\approx
\tau_0/2$ the signal to noise ratio will improve by a factor of two when
fourth differences are taken and an accuracy of about $0.1~\mu$Hz may be
sufficient to measure the characteristics of the signal. Such an accuracy
may be possible with the forthcoming space missions like COROT
\citep{baglin:98}, MOST \citep{matthews:98} and MONS \citep{kb:98}. The
oscillatory signal due to the \heii\ ionisation zone has an order of
magnitude larger amplitude for solar mass stars and it would be easier to
detect this signal. This signal may be useful in measuring the helium
abundance in stellar envelopes. The oscillatory signal in splitting
coefficients has much smaller amplitude and unless the stars are rotating
much faster and have a much larger $\delta\Omega$ it may not be possible
to detect such a signal.

\section{Signal from the edge of convective core}
\label{sec:corecz}

\subsection{Signal in frequencies of high mass stars}

\begin{figure}
\centering
\resizebox{\figwidth}{!}{\includegraphics{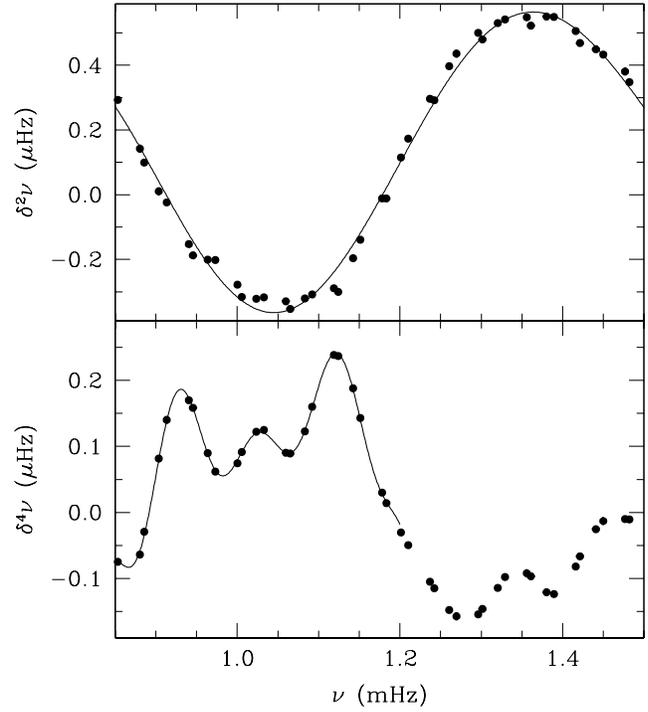}}
\caption{
	Second ({\it upper panel}) and fourth ({\it lower panel})
	differences of frequencies for a $2\msun$ star at an age of
	$0.5$~Gyr. Only modes with $\ell = 0$ to $3$ have been considered.
	Fits to the data with a function of the form described in
	Eq.~(\ref{eq:fourth_fit}) have been shown for the full range of
	frequencies in case of the second differences, and for a
	restricted range in case of the fourth difference. Note that
	the range of y-axis in this figure is an order of magnitude
	less than that in Fig.~\ref{fig:osc_fit}.
	}
\label{fig:sec_fourth}
\end{figure}

For stars with $M\ga1.2M_\odot$, apart from the outer convection zone, a
part of the inner core is also convective and we would expect an
additional oscillatory component in the frequencies arising from
discontinuity in derivatives of sound speed at the outer boundary of the
convective core. In this case, $\tau\approx\tau_0$ and as mentioned in the
previous section, taking fourth difference will reduce the amplitude
significantly. In order to understand the oscillatory signal, we consider
a stellar model for $M=2M_\odot$ at an age of $5\times10^8$ years. This
model has a convective core with acoustic radius $0.03\tau_0$ and taking
the fourth difference will reduce the amplitude by a factor of about
$1000$. The outer convection zone in this model has an acoustic depth of
about $0.08\tau_0$ and that oscillatory signal will also reduce by a
factor of $20$. Fig.~\ref{fig:sec_fourth} shows the second and fourth
differences of frequencies for this model. It is clear from the figure
that the oscillatory signal arising from the outer boundary of the
convective core is not seen in either the second or the fourth
differences. These differences do show the signal from the base of the
outer convective zone or the \heii\ ionisation region, while the
oscillations that are seen in the fourth differences have completely
different $\tau$. Fitting an oscillatory signal in a restricted frequency
range shows that this signal is arising from $\tau\approx 5300$~s, which
corresponds to $r/R\approx 0.6$, where we do not expect any boundary of
convective region.

\begin{figure}
\centering
\resizebox{\figwidth}{!}{\includegraphics{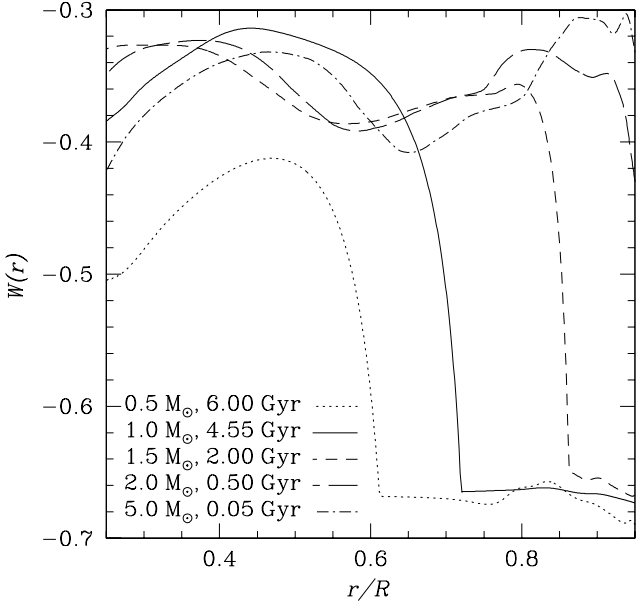}}
\caption{
	The quantity $W=(1/g)\D c^2/\D r$ for stars of
	different masses and ages.
}
\label{fig:wfunc}
\end{figure}

In order to understand this signal we show in Fig.~\ref{fig:wfunc}, the
quantity $W(r)=(1/g)\D c^2/\D r$ for a few different stellar models. It can
be seen that while low mass stars do not have any sharp variation in this
scaled derivative of the sound speed in the radiative interior, the high
mass stars all show a prominent dip in $W(r)$ in the radiative interior.
This dip is quite small compared to the sharp rise near the base of the
outer convection zone in near-solar mass stars. But for high mass stars
where the outer convection zone is very shallow, this dip becomes the
dominant feature in the first derivative of the sound speed and hence
contributes to the oscillatory signature. Moreover, calculating the fourth
difference increases the amplitude of this signal by more than an order of
magnitude, while amplitudes of other oscillatory terms coming from edges
of convective layers are significantly reduced. Thus in the fourth
difference the prominent oscillations are due to this dip in $W(r)$, while
in the second difference the oscillations arising from the base of the
outer convection zone are seen. But the expected signal from the outer
edge of the convective core is not seen in either the second or the fourth
differences. Since there is no discontinuity in the derivatives of the
sound speed in the radiative region the oscillatory signal does not have
the same form as that expected from a discontinuity, but for low
frequency modes, where the radial wavelength would be larger than the size
of the dip the signal is seen more clearly. At high frequencies the radial
wavelength may be comparable to or smaller than the size of the dip in
$W(r)$ and the oscillatory signal is not clear. Further, just like the
\heii\ ionisation zone, the amplitude of this oscillatory signal should
also be modulated by an oscillatory factor, which is not included in
our fits.

\subsection{Signal in splitting coefficients}

In order to understand our failure to see any oscillatory signal from the
convective core in the frequencies, we attempt to study the oscillatory
signal due to the tachocline in the splitting coefficients. We calculate
the splitting coefficients assuming the tachocline to have different
depths to study the variation in amplitude with depth and the results are
shown in Table~\ref{tab:ampl_depth}. We have considered a stellar model
with mass $2\msun$ and age $0.5$~Gyr for the purpose of illustration.
The amplitude appears to be maximum when the tachocline is located near
the acoustic midpoint of the star (in this particular case, near $r/R =
0.8$) and falls off as it shifts to either end. It can be easily seen that
this variation is more or less fully explained by the amplification factor
while taking the differences. Thus it appears that the amplitude of the
signal in the splitting coefficients is more or less independent of the
depth and the apparent variations are coming from calculating the
differences. It is clear that in order to detect the oscillatory signal
from the outer edge of the convective core, we should look at the
splitting coefficients directly, rather than their differences.
\begin{table}
\centering
\caption{
	The variation of amplitude of $\delta^{2}c_{3}$ with position of
	tachocline for a $2\msun$ star of age $0.5$~Gyr. All the
	amplitudes are calculated at the scaled frequency $1.19$~mHz,
	corresponding to $2.5$~mHz for a solar model. The total acoustic
	depth of the star is $\tau_0 = 8128$~s.
}
\label{tab:ampl_depth}
\begin{tabular}{cccc}
\hline
$r_{d}/R$ & $\tau_{d}$ & $A$ & $A/4\sin^2(\pi\tau_{d}/\tau_0)$ \\
\hline
 & (s) & (nHz) & (nHz) \\
\hline
0.30 & 7267 & 0.0540 & 0.1288 \\
0.50 & 6347 & 0.2058 & 0.1284 \\
0.70 & 4986 & 0.4758 & 0.1366 \\
0.80 & 4022 & 0.5813 & 0.1454 \\
0.90 & 2725 & 0.4385 & 0.1449 \\
0.95 & 1820 & 0.2482 & 0.1413 \\
\hline
\end{tabular}
\end{table}

\begin{figure}
\centering
\resizebox{\figwidth}{!}{\includegraphics{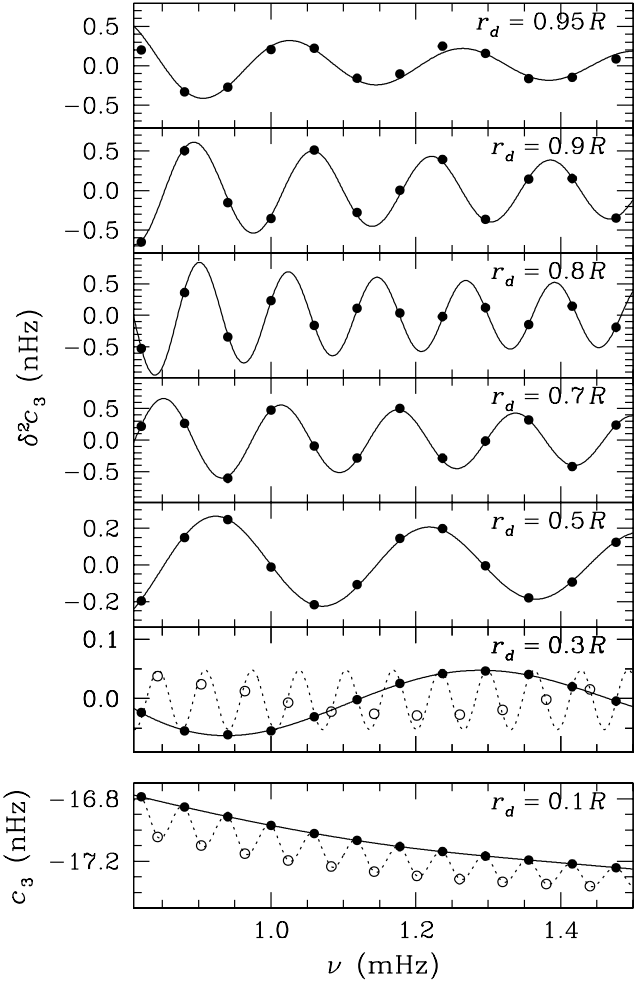}}
\caption{
	Second difference of the splitting coefficient $c_3$ for $\ell =
	2$ only ({\it filled circles}), for a $2\msun$ star at an age of
	$0.5$~Gyr, arising out of a tachocline at different depths,
	$r_{d}$ as marked in each panel.
	A fit to the data using a function of the form described
	in Eq.~(\ref{eq:fourth_fit}) is shown in each case. The two
	panels at the bottom shows the fit with the ``true'' value of 
	$\tau_d$ ({\it
	dotted line}), which is obtained when both $\ell=2$ and $3$ ({\it
	empty circles}) are considered. Compare this to a fit with the
	``aliased'' $\tau_d$ ({\it solid line}), obtained with $\ell=2$
	modes only.
}
\label{fig:fit_alias}
\end{figure}

Fig.~\ref{fig:fit_alias} shows the second difference of splitting
coefficients for $\ell=2$ modes arising from the tachocline at different
depths in the stellar model mentioned above. It can be seen that the
wavelength of oscillations reduces starting from $r_d=0.95R$ as we go
deeper as expected. However, after $r_d=0.8R$ the wavelength appears to
increase again contrary to what we expect. This is similar to the
phenomenon of aliasing in Fourier transform. For simplicity if we assume
that the frequencies of $\ell=2$ modes are uniformly spaced then the
oscillatory term can be written as
\begin{eqnarray}
c_3(n,\ell) & = & A\sin[4\pi\tau(\nu_0+n\Delta\nu_\ell)+\phi] \nonumber\\
            & = & A\sin[4\pi\tau\nu_0+\phi+2\pi n\tau/\tau_0],
\label{eq:alias}
\end{eqnarray}
where $\Delta\nu_\ell$ is the spacing between successive modes for given
$\ell$. Here we have used the fact that for low $\ell$,
$\Delta\nu_\ell\approx 1/(2\tau_0)$. Now if $\tau=\tau_0$ it is clear that
for all $n$ the oscillatory term will be the same and $c_3(n,\ell)$ will
not show any oscillations. If we take $\tau'=\tau_0-\tau$, then it is
clear that the oscillatory term will be the same apart from a change in
the phase for all modes. Thus from the oscillatory signal for an
individual $\ell$ it is not possible to distinguish between these two
values of $\tau$. Fig.~\ref{fig:fit_alias} shows the actual fit obtained
using two different $\tau$ for some depths. This is exactly similar to
aliasing in discrete Fourier transform. Alternatively, it means that we
can measure the acoustic distance from either boundary. If we consider
sampling at a uniform spacing of $\Delta\nu \approx 1/(2\tau_0)$, then the
corresponding Nyquist frequency will be $1/(2\Delta\nu)\approx \tau_0$.
Thus `frequencies' higher than $\tau_0$ will be aliased to lower
frequencies. Because of the factor of $2$ inside the sine function in
Eq.~(\ref{eq:fourth_fit}) this actually corresponds to an acoustic depth
of $\tau_0/2$. Thus this is identical to aliasing in discrete Fourier
transform.

It may be noted that the lowest panel in Fig.~\ref{fig:fit_alias} for
$r_d=0.1R$ is shown separately, as in this case instead of the second
differences (which have very small amplitude), we show the splitting
coefficients themselves. Although, the amplitude of variation in this is
comparable to amplitudes at other depths, it is impossible to distinguish
this signal from a smooth trend. At other depths we have shown the second
difference as the presence of smooth trend in splitting coefficients does
not show the oscillatory signal clearly. It is clear from this panel that
both $\ell=2$ and $3$ modes separately fall on a smooth curve and only if
they are combined together the oscillatory signal from the core can be
identified. It may be noted that the form of rotation rate used to
calculate these splitting coefficients does not have any smooth component
and hence the smooth component in resulting splitting coefficients is
quite small, thus enabling us to see the signal in some form. In real
stars there will also be some smooth component which will be difficult to
isolate from the oscillatory part when the apparent $\tau$ is quite small.

In principle, the ambiguity in $\tau$ can be resolved by including
multiple $\ell$ values. The two panels at the lower end of
Fig.~\ref{fig:fit_alias} shows the splitting coefficients for $r_d=0.3R$
and $0.1R$ for $\ell=2,3$. It is clear that although modes corresponding
to each $\ell$ show a variation with large wavelength, the two do not
agree with each other in phase and if we attempt a fit using
Eq.~(\ref{eq:fourth_fit}), the fit will not be good when $\tau$
corresponds to the smaller value implied by individual $\ell$. Only an
oscillatory form with larger value of $\tau$ can fit all the modes. In
terms of discrete Fourier transform this can be understood as follows:
because of additional $\ell$ the sampling interval is reduced thus
increasing the Nyquist frequency, enabling us to determine higher
frequencies. However, the problem with fitting any form is that in all
cases there will be some smooth component in the splitting coefficient
arising from smooth variations in the rotation rate and it will be
difficult to separate out this component which has to be done for each
$\ell$ separately, as the smooth component will depend on $\ell$ and
frequency.
The problem arises
because in this case the oscillatory component also appears to be smooth
due to aliasing and it will be difficult to isolate it from the smooth
component as can be seen by the panel for $r_d=0.1R$ in
Fig.~\ref{fig:fit_alias}. This separation will be more difficult when
errors are added to the splitting coefficients, as will be the case in
real observed values. Thus it is difficult to detect any possible
oscillatory signature arising from convective cores in high mass stars.

\subsection{Signal from the convective core}

\begin{figure}
\centering
\resizebox{\figwidth}{!}{\includegraphics{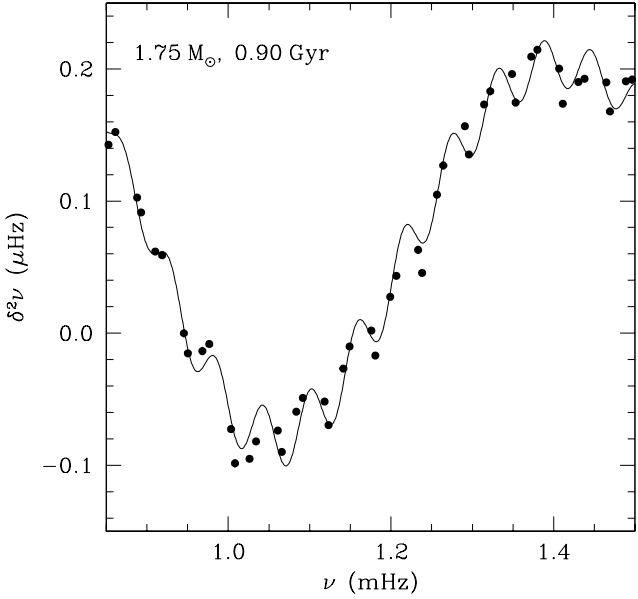}}
\caption{
	Second difference of the frequencies, for a $1.75\msun$ star at an
	age of $0.9$~Gyr. A fit to the data using a function having two
	similar components as described in Eq.~(\ref{eq:fourth_fit}) is
	shown. The high frequency oscillations are due to outer boundary
	of the convective core, while the dominant oscillation is due to a
	combination of \heii\ ionisation zone and the base of the convective
	envelope.
}
\label{fig:core_1.75}
\end{figure}

Detecting the oscillatory signal in the frequency from the convective core
is even more difficult as the frequencies have dominant oscillatory
contributions arising from other variations in sound speed derivatives.
Thus the signal due to the outer convection zone will have similar $\tau$
to that from convective core (after aliasing) and it will be difficult to
isolate the core contribution from the surface contribution. Further,
there is a smooth variation with $n$ for the frequencies, which is several
orders of magnitude larger than the oscillatory part and it will be more
difficult to remove this smooth variation. Of all the models tried we could
detect this signal in only $1.75M_\odot$ star with an age around $0.9$~Gyr
(Fig.~\ref{fig:core_1.75}). Interestingly even with exact frequency data
this signal is not clearly seen at all ages. We believe the main reason
for this is that at this particular age the amplitude of signal arising
from the outer convection zone and \heii\ ionisation zone combine to yield a
small amplitude, possibly because of beating between the two frequencies
which are reasonably close. Because of the small amplitude, even smaller
oscillatory signal due to the convective core shows up with amplitude of
order of $0.01~\mu$Hz, which would otherwise have been missed. This signal
will translate to an amplitude of $0.3~\mu$Hz in frequencies, which can be
compared with frequency separation of $50~\mu$Hz. Although, the amplitude
of this oscillatory signal is fairly large in the frequencies themselves
it gets reduced when differences are taken. If we do not take differences
we need to remove the smooth part independently, which is also difficult
as explained earlier. We also tried to use the first difference of
frequency instead of second difference, so that the amplitude of
oscillatory signal from the convective core would be larger. But even in
that case the smooth part is substantial and it is not easy to remove it
in most cases, thus making it difficult to identify the oscillatory
signal. 

In an attempt to detect the oscillatory signature we consider even higher
mass stars where the convective core is comparatively larger. Thus we
consider a stellar model with $M=5M_\odot$ and age $5\times10^7$ years. In
this case the convective core has a radius of $0.04\tau_0$ and
Fig.~\ref{fig:core_1.75} shows the second differences of the frequencies.
It appears that at a somewhat high frequency range some oscillatory signal
which is similar to what one would expect from the convective core is
present. However, the acoustic cutoff frequency for this model is only
$0.72$~mHz and hence the oscillatory pattern is seen only in high
frequency modes which are not trapped in the interior. These modes are
unlikely to be seen in distant stars. Even if they are seen, it will be
difficult to determine their frequencies accurately, because of large
widths in the power spectrum that may be expected. Similar signal at
frequencies above the acoustic cutoff has been seen in some other stellar
models also. The amplitude of dominant oscillations due to the \heii\ 
ionisation zone falls off rapidly with frequency and that probably allows
the smaller signal arising from the convective core to be seen at high
frequencies.
\begin{figure}
\centering
\resizebox{\figwidth}{!}{\includegraphics{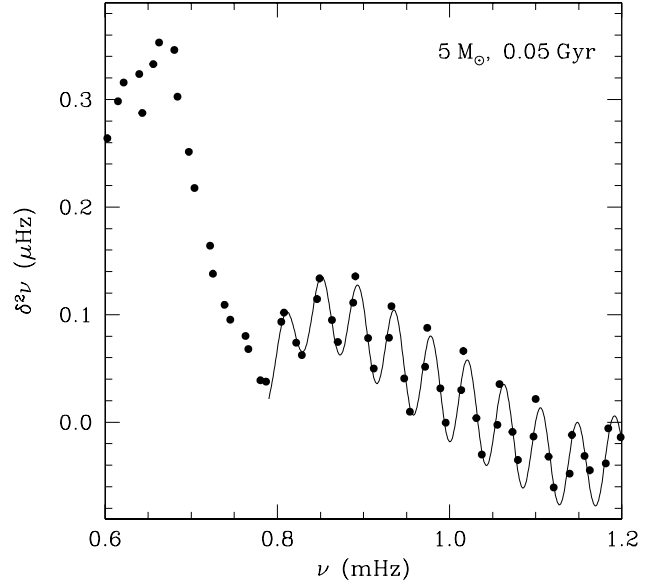}}
\caption{
	Second difference of the frequencies, for a $5\msun$ star at an
	age of $0.05$~Gyr. A fit to the data in a restricted frequency
	range using a function having two similar components as described
	in Eq.~(\ref{eq:fourth_fit}) is shown.
}
\label{fig:core_5.00}
\end{figure}

Since we have failed to detect the oscillatory signal from convective
core, in the next section we propose
an alternative technique which looks at the variations in the smooth part
to detect convective cores in stars and possibly measure its size.

\section{Signature of convective core in frequency separation}

From the asymptotic theory of stellar oscillations it is well known that
the frequency separations
\begin{eqnarray}
\Delta \nu_\ell & = & \nu_{n+1,\ell}-\nu_{n,\ell}, \nonumber \\
D_\ell & = & {\nu_{n,\ell}-\nu_{n-1,\ell+2}\over 4\ell+6}, \\
d_{1/2} & = & {3\over 2}(\nu_{n,0} - 2\nu_{n,1} + \nu_{n+1,0}), \nonumber 
\label{eq:freq_sep}
\end{eqnarray}
are related to variation in the sound speed across the stellar radius
including the convective core. Some of these frequency separations are
quite sensitive to conditions in stellar core. These frequency separations
can be easily measured for stars using only the low degree modes. Here we
consider only the frequency separations for $\ell=0$ averaged over a
frequency range scaled appropriately for each model to correspond to a
range of $3$--$4$~mHz for the Sun.

\begin{figure}
\centering
\resizebox{\figwidth}{!}{\includegraphics{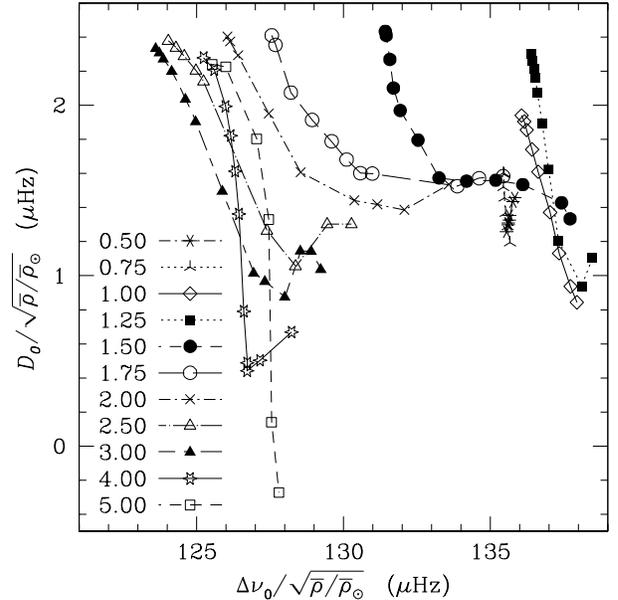}}
\caption{
	Evolutionary tracks for different stellar masses in the
	($\Delta\nu_{0}$, $D_{0}$) diagram. Both the frequency separations
	have been scaled by $\rhos$.
}
\label{fig:nu_d0}
\end{figure}
\citet{jcd:93} has suggested that these frequency separations can be
related to the stellar age and mass from the evolutionary tracks in the
($\Delta\nu_0$,$D_0$) plane, which is also known as the C-D diagram. Since
the frequencies of stellar oscillations are expected to scale as
$\sqrt{M/R^3}$ it will be more instructive to scale these frequency
differences by this factor to isolate other effects in the frequency
variation. Fig.~\ref{fig:nu_d0} shows the scaled frequency separations
plotted against each other for various stellar models. This is similar to
the figure shown by \citet{jcd:98}. It can be immediately seen that after
scaling the variation in the frequency separation is significantly reduced
and moreover, all models for $M\la 1.25M_\odot$ appear to have the scaled
large frequency separation $\Delta\nu_0/\rhos\approx 137~\mu$Hz, close to
the solar value. These are the models that do not have any significant
convective core. All models with convective core appear to have
significantly smaller scaled frequency separations. Thus it appears that
if the stellar mass and radius are known then by looking at the scaled
frequency separation we can infer the presence of a convective core
\citep[cf.,][]{thompson:00}. If the scaled large frequency separation is 
less than about $134~\mu$Hz, then the star is likely to have a convective 
core. 

\begin{figure}
\centering
\resizebox{\figwidth}{!}{\includegraphics{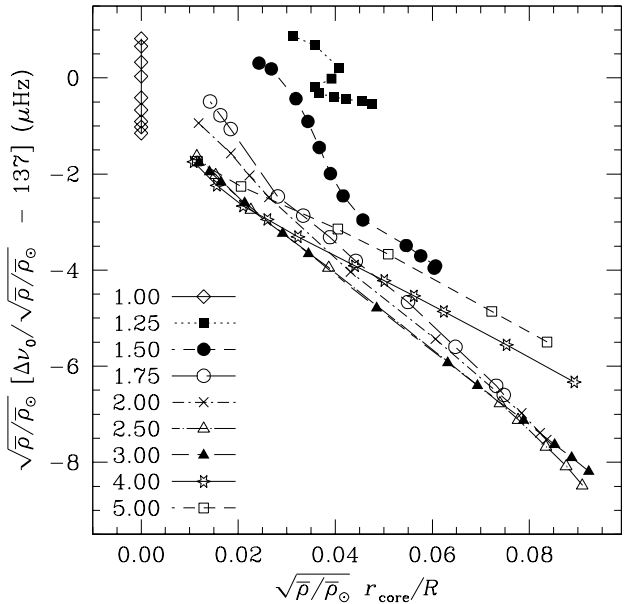}}
\caption{
	Large frequency separation, $\Delta \nu_{0}$, as a function of 
	convective core radius. Both the axes 
	have been scaled by $\rhos$.
}
\label{fig:rc_nu}
\end{figure}

Apart from detecting the presence of convective core we would also like to
measure its size. Indeed, the asymptotic theory of stellar oscillations
\citep[e.g.,][]{rv:01} may be able to provide a relation connecting the
frequency separation to the size of the convective core. In this work,
using the frequencies computed from different stellar models we attempt to
check if the deviation from the solar value of the large frequency
separation can be correlated to the size of the convective core. It might
also be possible to derive some combination of the large and small
frequency separations which would vary in tandem with the size of the
convective core. In Fig.~\ref{fig:rc_nu} we have plotted the difference of
the scaled large frequency separation, from the solar value of
$137~\mu$Hz, against the radius of the convective core for stellar models
of different mass and age. Both the axes have been scaled to bring the
curves corresponding to different masses close to each other. We find that
there is an approximately linear relation between these two quantities for
a wide range of masses. Although, all the curves do not collapse into one
single relationship, this diagram might be used to estimate the size of
the convective core, provided the mass and radius of the star are
independently known. Only the curves for $1.25M_\odot$ and $1.5M_\odot$
stars appear to fall outside the band.
\begin{figure*}
\centering
\hbox to \hsize{\hfil
\resizebox{\dblfigwidth}{!}{\includegraphics{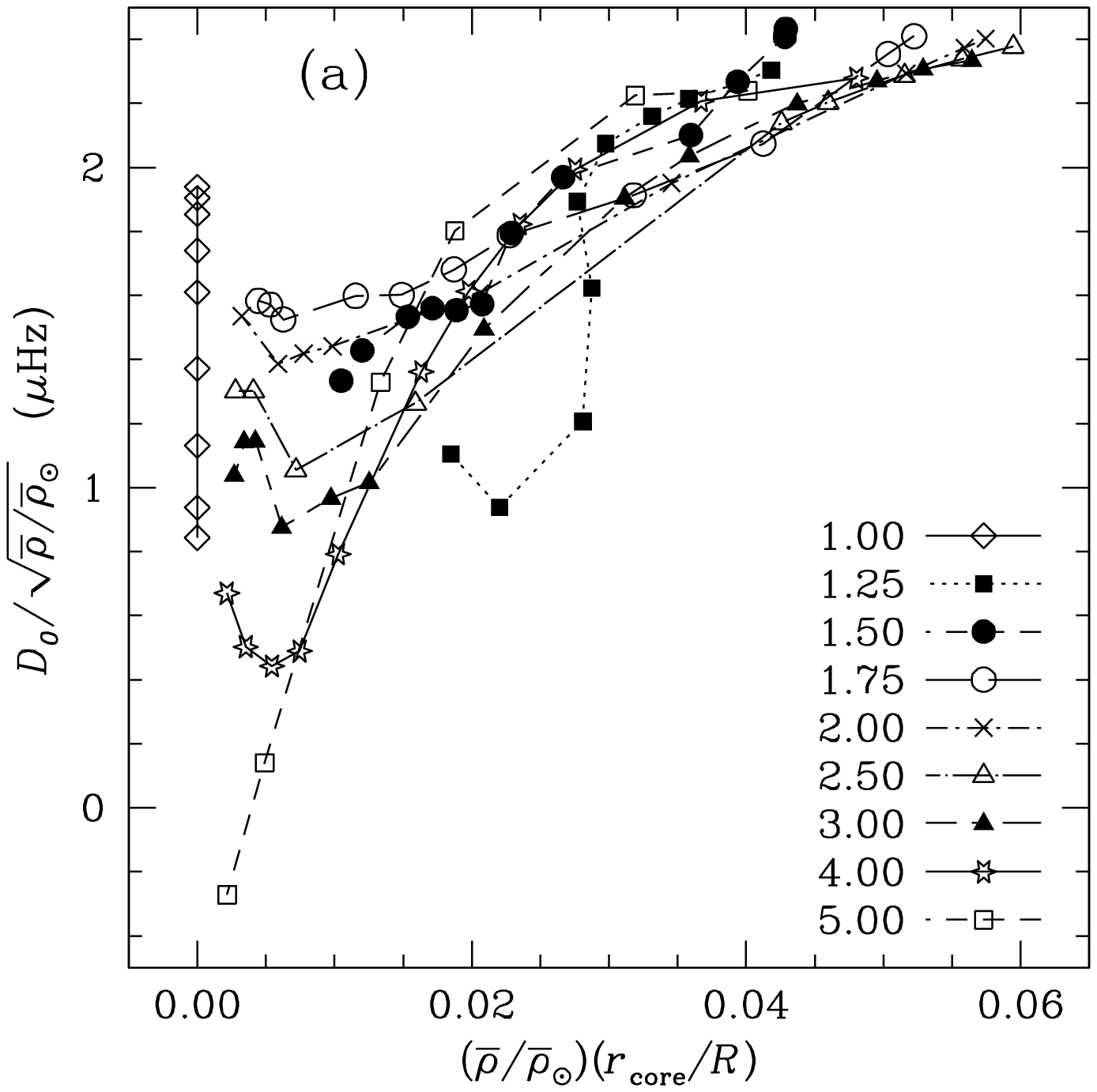}}
\hfil
\resizebox{\dblfigwidth}{!}{\includegraphics{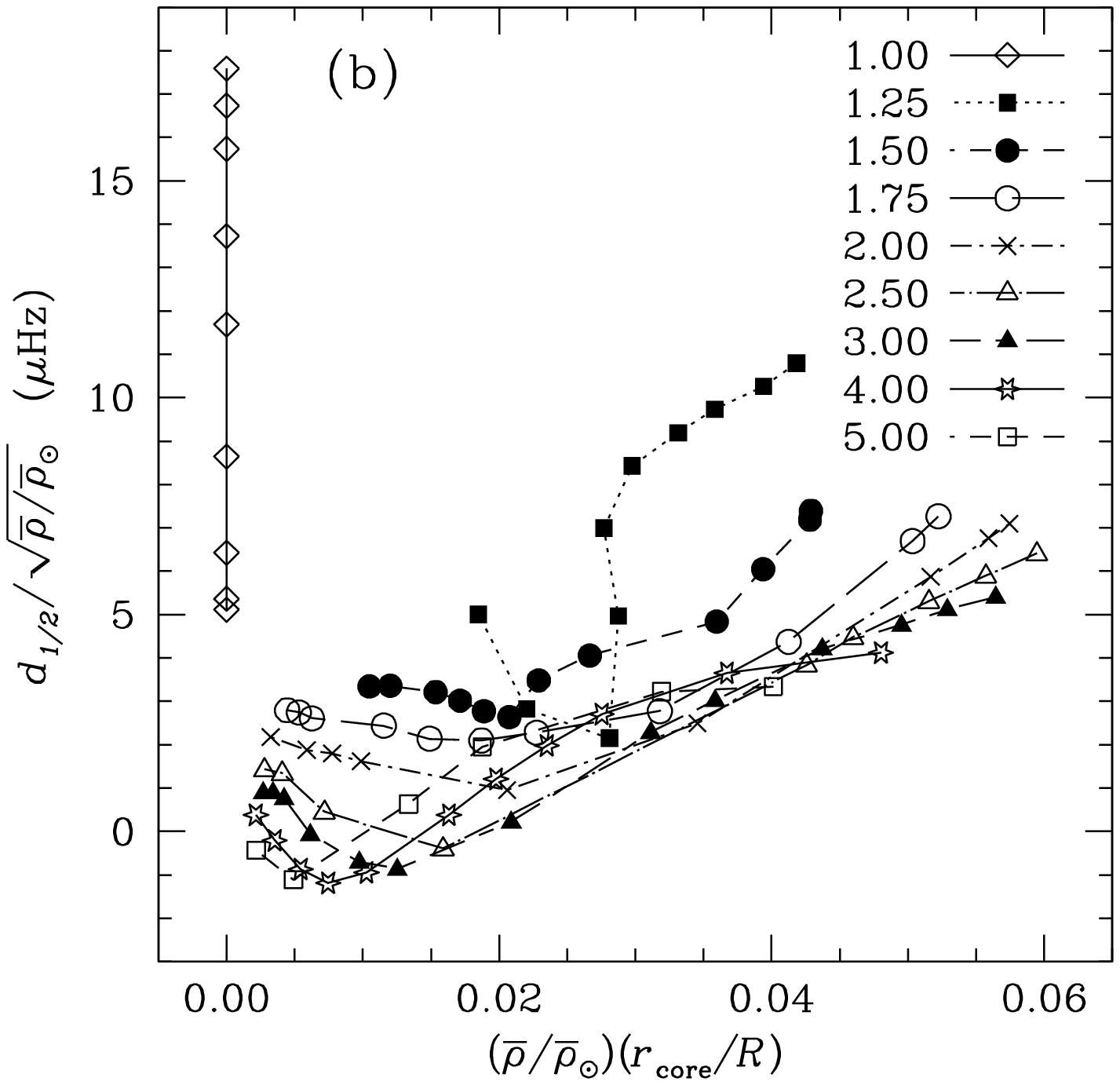}}
\hfil}
\caption{
	Small frequency separations, $D_{0}$ (in {\bf (a)}), 
	and $d_{1/2}$ (in {\bf (b)}) as functions of 
	convective core radius. Both the axes 
	have been scaled by $\rhos$.
}
\label{fig:rc_ss}
\end{figure*}

The small frequency separation, $D_0$, is found to bear a correlation with
the radial size of the convective core. This relation becomes clear with
appropriate scaling of $D_{0}$ as well as $r_{\mathrm{core}}$, as shown in
Fig.~\ref{fig:rc_ss}(a). One problem with using $D_0$ to measure the
convective core size is that the range of values covered by a solar mass
star during its evolution has a large overlap with the value for stars
with convective core. Thus by looking at the value of $D_0$ alone it will
be difficult to determine the convective core size.

\citet{basu:01} have pointed out that the behaviour of the small
separation, $d_{1/2}$, is somewhat different from the other frequency
separations. We find that this quantity does not have a clear relationship
with the size of the core. Nevertheless, as shown in
Fig.~\ref{fig:rc_ss}(b), its value derived for models with a convective
core (after some scaling) is conspicuously different from that of models
without convective core. However, for both the small separations, we are
unable to collapse all the curves together as would have been desirable
for them to be used as definitive diagnostics for the size of a convective
core. However, they can be used as complementary diagnostics for
properties of convective core.

An advantage of using the frequency separation as opposed to the
oscillatory signal discussed in the previous section is that comparatively
lower accuracy is required in this case. Thus an accuracy of order of
$1~\mu$Hz may be sufficient to get an estimate of convective core size
using the frequency separations. On the other hand, extracting the
oscillatory signal will require at least an order of magnitude higher
accuracy in measured frequencies. Even with very high accuracy it will be
difficult to isolate the oscillatory signal from convective core from
other signals. However, the use of frequency separation requires the
knowledge of stellar mass and radius for scaling the frequencies. This may
not be a problem for nearby stars which are likely to be the prime targets
of asteroseismic missions.

\section{Conclusions}

In this work we have tried to study the oscillatory signal in the
frequency arising from discontinuities in derivatives of sound speed
inside a star. This oscillatory signal can be conveniently used to study
the location of the base of the outer convection zone or the \heii\ 
ionisation zone as has been suggested by \citet{mct:00}. Our results also
support their conclusions. We have studied this signal in a wide range of
stellar models covering a range of stellar masses and ages. One of the
reasons for this study was to possibly identify promising stars where such
signal may be large enough to be detected. However, from our study it
appears that there is only a small variation in the amplitude of the
oscillatory signal arising from the base of the convection zone between
stars of different masses and ages on the main sequence.
The amplitude increases with decreasing mass or age.
The amplitude of
the oscillatory signal arising from the \heii\ ionisation zone appears to
increase with stellar mass and can possibly be used to measure helium
abundance in stellar envelopes. Typical amplitude of oscillations from the
base of the outer convection zone is around $0.05~\mu$Hz and from the
simulations that we have carried out it appears that comparable accuracy
will be required to extract this signal and measure the characteristics of
the convection zone boundary. Although, in this study, we have not included
any overshoot below the convection zone, it is well known that the
amplitude of this signal increases with overshoot and it will be possible
to use this signal to measure the extent of overshoot below the base of
the convection zone \citep{mct:00}. For the case of the Sun, this has
already been done, though with the help of higher degree modes whose
frequencies are known more accurately. The oscillatory signal arising from
the \heii\ ionisation zone is an order of magnitude larger and it should be
possible to detect this signal in observed frequencies.

In higher mass stars we find an oscillatory signal arising
from the dip in $W(r)$ in the radiative interior,
which may also provide some diagnostic for
stellar models. From Fig.~\ref{fig:wfunc} it can be seen that the location
of the dip depends on stellar mass. Since this location can be determined
through $\tau$ in oscillatory signal it can provide some constraints on
stellar models. It can be shown that in radiative regions $W(r)$ is also
affected by opacity, which in turn will depend on heavy element abundance,
$Z$. Thus the depth of the dip also depends on $Z$. However, the amplitude
of oscillations arising from this dip is about $0.05~\mu$Hz in the fourth
differences and hence we will need accuracy of order of $0.01~\mu$Hz
in measured frequencies to
detect this signal. This level of accuracy may not be achieved in
immediate future, but with further improvements and longer observations it
may be possible to detect this signal.

Most of the earlier studies have concentrated on the oscillatory signal
from the outer convection zone. It is clear that there would be a
discontinuity in the derivative of sound speed at the outer edge of the
convective core in a massive star. This discontinuity should also give
rise to a similar oscillatory signal. Although, \citet{mct:98} have
suggested that this signal can be used to study characteristics of
convective core, it is not clear if this signal has been seen in
frequencies of a stellar model. Thus in this work we attempt to study the
oscillatory signature arising from a convective core. Even using exact
frequencies from stellar models we find it difficult to detect this
signal. We have tried to identify the reasons for this failure by studying
a similar signal in the splitting coefficients arising from a possible
tachocline at different depths. We find that the oscillations display a
phenomenon similar to aliasing in discrete Fourier transform, because of
which using modes from a single value of $\ell$, it is not possible to
distinguish between oscillations arising from discontinuities at acoustic
depths of $\tau$ and $\tau_0-\tau$, where $\tau_0$ is the total acoustic
radius of the star. Due to this aliasing the signal from the convective
core would appear as a smooth variation in frequencies from any individual
$\ell$. Though, in principle, by combining data for $\ell=0,1,2,3$ we
should be able to characterise this signal, but in practice we have to
remove the smooth part for each $\ell$ before the oscillatory part can be
studied. It is not easy to separate this signal from the convective core,
thus making it difficult to study its characteristics.

From our attempts to find the oscillatory signal from the convective core,
we could find this signature in a few stellar models at frequencies above
the acoustic cutoff frequencies. Such modes are not likely to be observed
in stars, but from these fits it appears that the amplitude of the
oscillatory signal in frequencies is actually quite large, around
$0.3~\mu$Hz, but because of our inability to separate it from the smooth
component it is difficult to detect this signal. Of all the models that we
tried we could fit this signal only in a $1.75M_\odot$ star around the age
of $0.9$~Gyr at frequencies less than the acoustic cutoff frequency. This
is most probably because the dominant oscillatory signal due to outer
convection zone and \heii\ ionisation zone combine to yield a small
amplitude in this stellar model, because of beating.
Thus we conclude that an oscillatory
signal of the form studied in this work does not provide a viable means of
studying convective cores in massive stars.

Having failed to detect the oscillatory signature from the convective core
in frequencies of stellar oscillations, we attempt to check if the smooth
part of the frequency can be used to study the properties of the core. The
frequency separations $\Delta\nu_\ell$, $D_\ell$ and $d_{1/2}$ defined by
Eq.~(\ref{eq:freq_sep}) are known to be sensitive to stellar structure
including the core \citep[e.g., ][]{jcd:93,basu:01}. Since the frequencies
of stellar oscillations generally scale with $\sqrt{M/R^3}$ it is
instructive to consider the scaled frequency separations. It turns out
that the scaled large frequency separation is generally around $137~\mu$Hz
for stars without convective core, while for massive stars with convective
cores it tends to be significantly smaller. Thus this scaled frequency
separation can be used to detect the presence of a convective core. After
suitable scaling the departure of this frequency separation from the mean
value for stars without a convective core can be roughly correlated to the
size of the convective core as can be seen from Fig.~\ref{fig:rc_nu}. It
can be seen that apart from intermediate masses of $1.25M_\odot$ and
$1.5M_\odot$, for higher masses all the evolutionary tracks lie in a band
and from the measure of the frequency separation together with the scaling
factor it should be possible to estimate an approximate size of the
convective core. The small frequency separations $D_0$ and $d_{1/2}$ are
also sensitive to properties of the core, though we could not find a
similar tight relation with the core size. We believe that by combining
all the information in the smooth component of the frequency for only low
degree modes it should be possible to study some properties of the
convective core, including its size.
An advantage of using these
frequency separations is that comparatively lower accuracy of $1~\mu$Hz in
frequency would be sufficient, while we would need much higher accuracy of
about $0.05~\mu$Hz to study the oscillatory signal. Even the oscillatory
signal from the base of the solar convection zone has been mostly studied
using intermediate degree modes which have higher accuracy.
 In this work we have considered
only the mean frequency separations which naturally discards some
information, which is contained in the individual frequency separations
for each value of $n$. In particular, the variation of frequency
separation with $n$ will also give some useful information about
the convective core.
Instead of frequency separation
inversion techniques using only low degree modes have also been tried to
study stellar cores \citep{basu:01}.

In the foregoing discussion it is assumed that the scaling factor
$\sqrt{M/R^3}$ is known for stars. This would require knowledge of stellar
mass and radius. In principle, the stellar mass and age can be determined
from the C-D diagram \citep{jcd:93}, while if the distance to the star is
known, its luminosity and hence the radius can be determined. From
Fig.~\ref{fig:nu_d0} it can be seen that the scaled large frequency
separation is not very sensitive to stellar mass or age. The variation
over the entire range of stellar models considered by us is less than
10\%. Hence in absence of any other information the measured large
frequency separation can also be used to measure the scale factor. But in
that case the resulting separation cannot be used to measure the size of
convective core.

\section*{Acknowledgements}

We thank P. Morel for providing the CESAM code for stellar evolution and
for the help in its usage, and S. M. Chitre for useful discussions.
We also thank the Referee, M. J. P. F. G. Monteiro for useful suggestions.

\end{document}